\documentclass[]{elsart1}

\usepackage{epsfig}
\usepackage{graphics}
\usepackage{graphicx}
\usepackage{url}
\usepackage{subfigure}

\newtheorem{proposition}{Proposition}
\newtheorem{finding}{Finding}

\pdfoutput=1

\begin{document}

\begin{frontmatter}

\title{Slime mould logical gates:\\ exploring ballistic approach}

\author{Andrew Adamatzky}

\address{University of the West of England, Bristol, United Kingdom\\ {\tt andrew.adamatzky@uwe.ac.uk}}

\date{\today}

\begin{abstract}
Plasmodium of \emph{Physarum polycephalum} is a single cell visible by unaided eye. 
On a non-nutrient substrate the plasmodium propagates as a traveling 
localization, as a compact wave-fragment of protoplasm. The plasmodium-localization 
travels in its originally predetermined direction for a substantial period of time even when 
no gradient of chemo-attractants is present. We utilize this property of \emph{Physarum}
localizations to design a two-input two-output Boolean logic gates 
 $\langle  x, y \rangle \rightarrow \langle  xy, x+y \rangle$ and 
 $\langle  x, y \rangle \rightarrow \langle  x, \overline{x}y \rangle$. 
We verify the designs in laboratory experiments and computer simulations. 
We cascade the logical gates into one-bit half-adder and simulate its 
functionality. 

\vspace{0.5cm}

\end{abstract}


\maketitle

\end{frontmatter}

\section{Introduction}

Plasmodium of \emph{Physarum polycephalum} is a single cell with many diploid nuclei. The cell is visible by naked eye and can grow up to meters when properly cared for. The plasmodium feeds on microscopic food particles, including microbial life forms. 
The plasmodium placed in an environment with distributed nutrients  develops a network of protoplasmic tubes spanning the nutrients' sources. 

In its foraging behavior the plasmodium approximates shortest path~\cite{nakagaki_2001a}, computes planar proximity 
graphs~\cite{adamatzky_toussaint} and plane tessellations~\cite{shirakawa}, exhibits primitive memory~\cite{saigusa}, 
realizes basic logical computing~\cite{tsuda_2004}, and controls robot navigation~\cite{tsuda_2007}. The plasmodium 
can be considered as a general-purpose computer because the plasmodium simulates Kolmogorov-Uspenskii machine --- the 
storage modification machine operating on a labeled set of graph nodes~\cite{adamatzky_ppl_2007}.  

In 2004 Tsuda, Aono and Gunji~\cite{tsuda_2004} demonstrated in laboratory experiments realisation of 
Boolean logic negation and conjuction by plasmodium of \emph{Physarum polycephalum}. In 2004 Adamatzky and De Lacy Costello esablished in numerical simulation~\cite{adamatzky_2004_collision} and chemical laboratory experiments~\cite{ben_2005} 
that by colliding localized excitations, or wave-fragments, in excitable chemical medium one can implement
functionally complete set of logical gates. We merge approaches~\cite{tsuda_2004} and ~\cite{adamatzky_2004_collision,ben_2005}
in present paper. We adapt concepts of collision-based computing~\cite{adamatzky_cbc} to realms of Physarum behaviour, and develop experimental prototypes of two-input two-output Boolean logical gates. 

The paper is structured as follows. Methods of cultivating and experimenting with plasmodium of \emph{Physarum polycephalum} 
are described in Sect.~\ref{methods}. In Sect.~\ref{ballistic} we provide experimental evidence of `ballistic' behavior of 
traveling plasmodium localizations. Experimental Physarum gates are discussed in Sect.~\ref{physarumgates}. 
In Sect.~\ref{simulation} experimental results are supported by numerical simulation of propagating localizations. The gates are cascaded in one-bit half-adder in Sect.~\ref{adder}. Importance of non-nutrient substrate for gate implementation is highlighted in Sect.~\ref{discussion}.

\section{Materials and Methods}
\label{methods}

Plasmodium of \emph{Physarum polycephalum} is cultivated in large plastic boxes, on a wet paper towels and fed with oat flakes.
Experiments are conducted in round Petri dishes (9~cm in diameter) and rectangular Petri dishes (12~cm $\times$ 12~cm).
Channels and junctions physically representing logical gates are cut of a non-nutrient 2\% agar plates
(Select agar, Sigma Aldrich). The dishes are kept in room temperature (c. $25^o$) in darkness. Images of plasmodium propagating in Petri dishes are taken by Epson Perfection 4490 scanner, resolution 600. Colors are enhanced by increasing saturation and contrast.

We use two-variable Oregonator model to numerically simulate propagation of plasmodium localizations. Our choice and 
details of the model are outlined below. 

Localized excitations in sub-excitable Belousov-Zhabotinsky (BZ) medium behave similarly to pseudopodia of \emph{P. polycephalum}~\cite{adamatzky_physarum_bz,adamatzky_bz_trees}. Sources of nutrients are chemo-attractants for plasmodium, 
gradients of shade are 'photo-attractants' for excitation waves in BZ medium. In~\cite{adamatzky_bz_trees} 
we shown how to navigate traveling localizations and growing parts of plasmodium by spatial configuration of 
attractants. We adopt the analogy developed in~\cite{adamatzky_bz_trees} and simulate propagating plasmodium 
using two-variable Oregonator equation~\cite{field_noyes_1974} adapted to a light-sensitive 
BZ reaction with applied illumination~\cite{beato_engel}:

$$\frac{\partial u}{\partial t} = \frac{1}{\epsilon} (u - u^2 - (f v + \phi)\frac{u-q}{u+q}) + D_u \nabla^2 u$$
$$\frac{\partial v}{\partial t} = u - v .$$

In framework of BZ reaction the variables $u$ and $v$ represent local concentrations of activator, or 
excitatory component, and inhibitor, or refractory component. With regards to plasmodium of \emph{P. polycephalum}
activator, $u$, is analogous to concentration, or `thickness', of the plasmodium's cytoplasm at the propagating 
pseudopodium. Inhibitor, $v$, combines several factors, when plasmodium is concerned. These factors include 
rate of nutrients consumption, byproducts of biochemical chains ignited by signals on photo- and 
chemo-receptors,  and concentration of metabolites released by the plasmodium into its substrate.

Parameter $\epsilon$ sets up a ratio of time scale of variables $u$ and $v$, $q$ is a scaling parameter 
depending on rates of activation/propagation and inhibition,  $f$ is a stoichiometric coefficient. 
Constant $\phi$ is a rate of inhibitor production. In light-sensitive BZ $\phi$ represents rate of inhibitor 
production proportional to intensity of illumination. In terms of plasmodium $\phi$ represents rate of inhibitor
proportional to concentration of nutrients, metabolites, illumination, chemical repellents. 
See detailed comparison of BZ and Physarum in~\cite{adamatzky_physarum_bz,adamatzky_bz_trees}. 
  
To integrate the system we use Euler method with five-node Laplacian operator, time step $\Delta t=0.001$ and 
grid point spacing $\Delta x = 0.25$ (equivalent to 0.6~mm of physical space), $\epsilon=0.0243$,  $f=1.4$, $q=0.002$.  The parameter $\phi$ characterizes excitability of the simulated medium: the medium is excitable and exhibits `classical' target waves when $\phi=0.05$ and  the medium is sub-excitable with propagating localizations, or wave-fragments, when $\phi=0.0766$.

\section{Ballistic of Physarum localizations}
\label{ballistic}

\begin{proposition}
Given cross-junction of agar channels and plasmodium inoculated in one of the channels, the plasmodium propagates straight through
the junction. 
\end{proposition}

\begin{figure}
\centering
\subfigure[]{\includegraphics[width=0.47\textwidth]{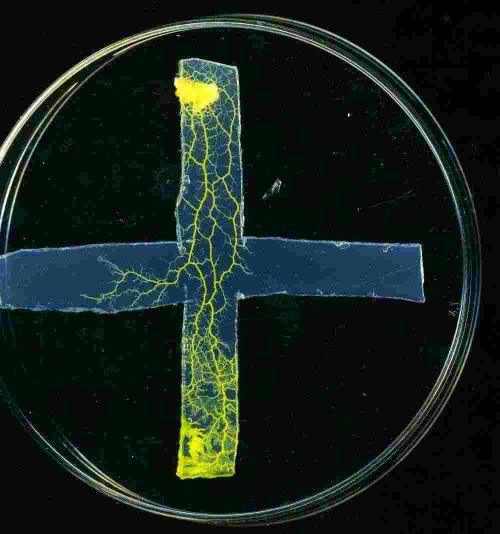}}\\
\subfigure[]{\includegraphics[width=0.45\textwidth]{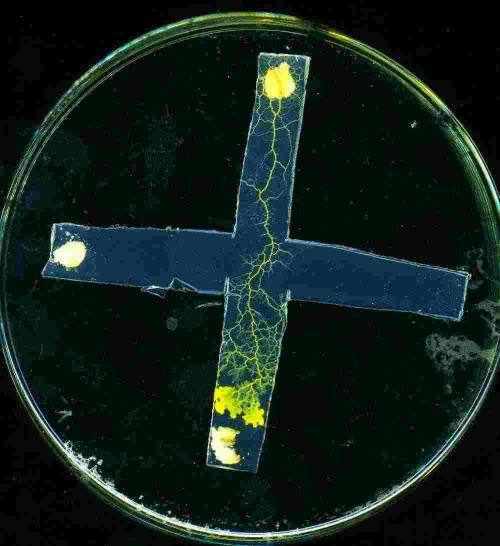}}
\subfigure[]{\includegraphics[width=0.47\textwidth]{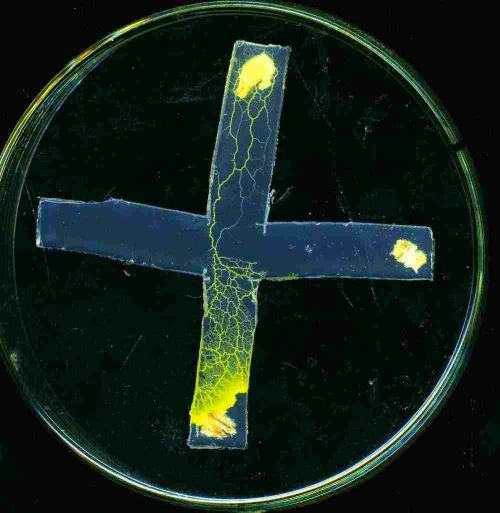}}
\caption{Experimental examples of plasmodia moving under their own momenta: 
(a)~no sources of chemo-attractants, 
(bc)~oat flakes are placed in south and east~(b), 
and south and west~(c) ends of channels.}
\label{cross}
\end{figure}

We experimentally found that that plasmodium propagates under its own momentum when no gradients of 
repellents or attractants are applied. In example shown in Fig.~\ref{cross}a the plasmodium is inoculated in 
northmost part of the north-south channel. The plasmodium has the only option --- to propagate south --- because
there is no gel substrate further north. Thus an internal `momentum' is formed. No food sources are applied 
on the substrate to attract the plasmodium. The plasmodium propagates straightforwardly by itself. 
It does not branch at the junction and moves till it reaches south end of the north-south channel (Fig.~\ref{cross}a). 

In 21 out of 28 trials the plasmodium exhibits clear `ballistic' behavior and propagates straight through the junction 
as under its own momentum. In 7 out of 28 trials the plasmodium turns into other channels or branches into several channels
at once. Adding sources of attractants --- assuming they are balanced over the channels (i.e. if there is an attractant in east-west channel there should be one in north-south channel) slightly speeds up propagation of plasmodium and does not change overall statistics of the plasmodium propagation. Examples are shown in Fig.~\ref{cross}b--d. Thus in experiments described in Sect.~\ref{physarumgates} we always used oat flakes to stimulate the plasmodium growth.

\section{Physarum gates}
\label{physarumgates}

\begin{figure}
\centering
\includegraphics[width=0.33\textwidth]{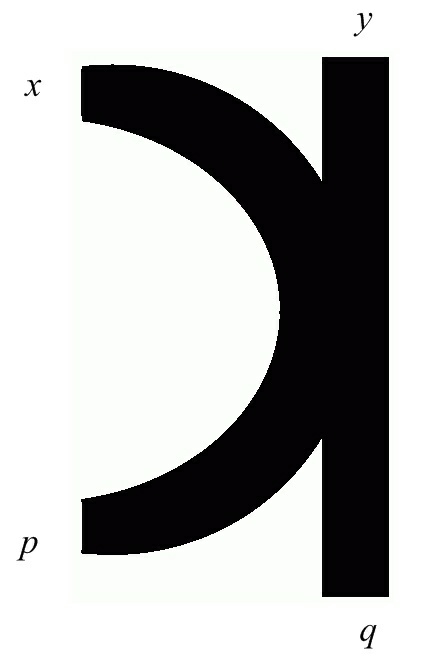} \hspace{1cm}
\includegraphics[width=0.43\textwidth]{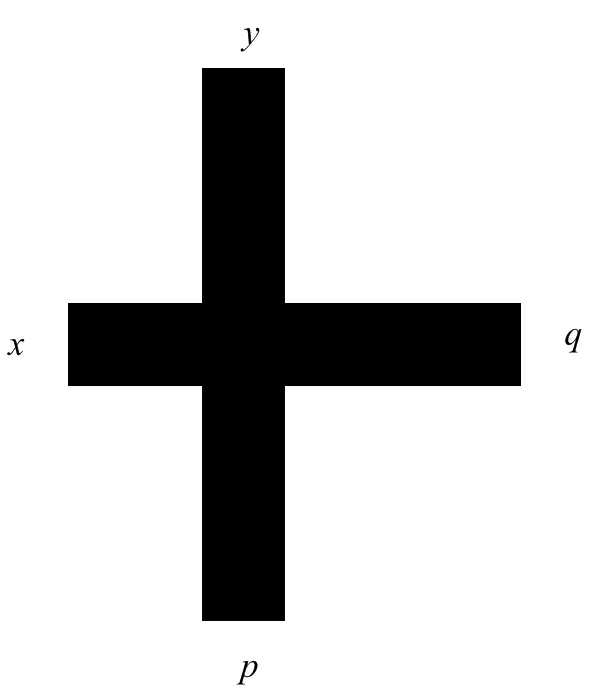}
\caption{Geometrical structure of Physarum gates $P_1$~(a) and $P_2$~(b): $x$ and $y$ are inputs, $p$ and $q$ are outputs.}
\label{gatescheme}
\end{figure}

Geometrical structure of gates $P_1$ and $P_2$ is shown in Fig.~\ref{gatescheme}. We experimented with various shapes of agar
and found that the most suitable templates are those shown in Fig.~\ref{gatescheme}. Input variables are $x$ and $y$ and 
outputs are $p$ and $q$. Presence of a plasmodium in a given channel indicates {\sc Truth} and 
absence --- {\sc False}.

\begin{figure}
\centering
\subfigure[]{\includegraphics[width=0.3\textwidth]{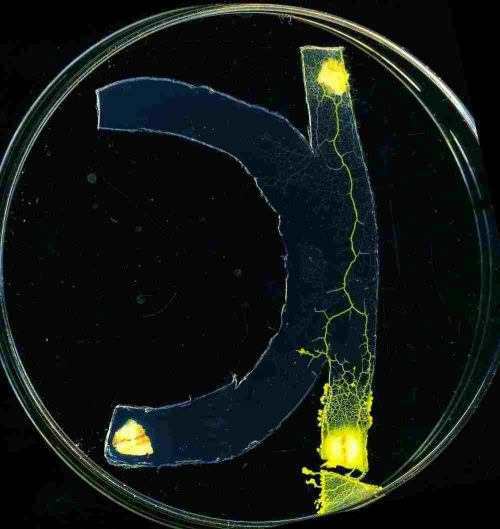}}
\subfigure[]{\includegraphics[width=0.3\textwidth]{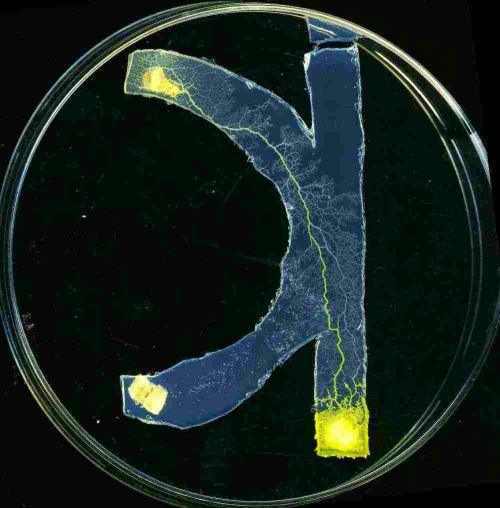}}
\subfigure[]{\includegraphics[width=0.3\textwidth]{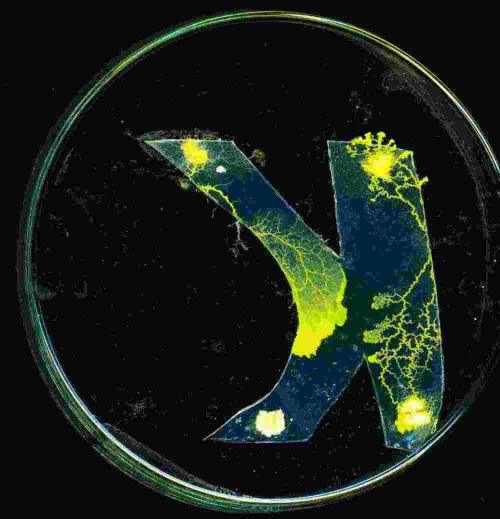}}
\subfigure[]{\includegraphics[width=0.3\textwidth]{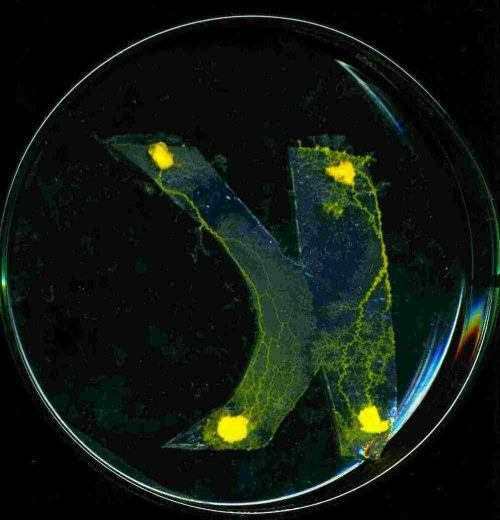}}
\subfigure[]{\includegraphics[width=0.3\textwidth]{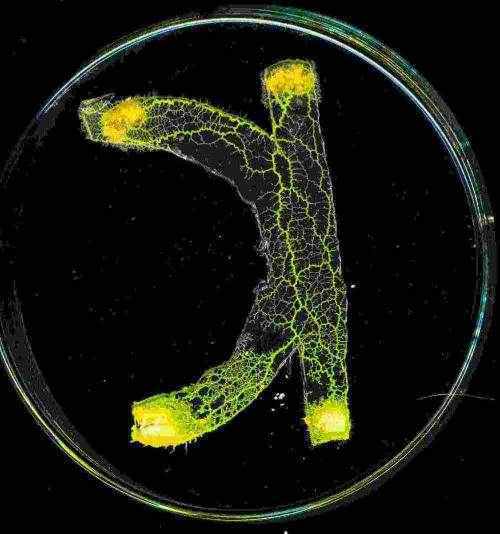}}
\subfigure[]{\includegraphics[width=0.3\textwidth]{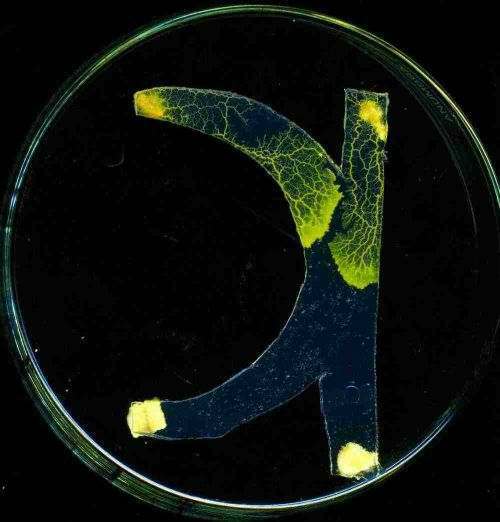}}
\subfigure[]{\includegraphics[width=0.3\textwidth]{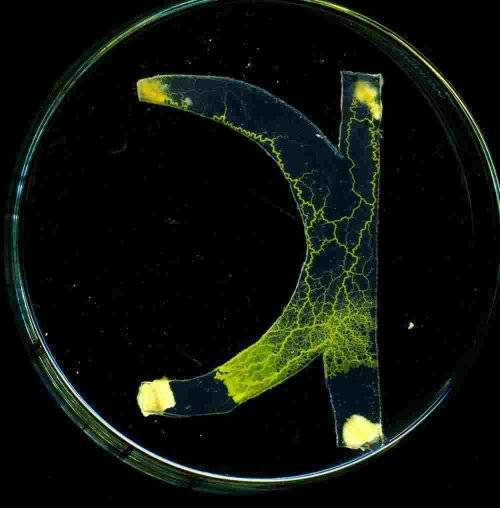}}
\caption{Experimental examples of 
transformation $\langle x, y \rangle \rightarrow \langle p, q \rangle$
implemented by Physarum gate $P_1$. 
(a)~ $\langle 0, 1 \rangle \rightarrow \langle 0, 1 \rangle$,
(b)~$\langle 1, 0 \rangle \rightarrow \langle 0, 1 \rangle$,
(c)--(e)~$\langle 1, 1 \rangle \rightarrow \langle 1, 1 \rangle$.
(f)--(g) snapshots of $P_1$ gate taken in 12~h interval.}
\label{gateinputs}
\end{figure}

Each gate implements a transformation from $\langle x, y \rangle \rightarrow \langle p, q \rangle$. 
Experimental examples of the transformations are shown in Fig.~\ref{gateinputs}. Plasmodium inoculated in input $y$ propagates
along the channel $yq$ and appears in the output $q$ (Fig.~\ref{gateinputs}a). Plasmodium inoculated in input $x$ propagates
till junction of $x$ and $y$, `collides' to the impassable edge of channel $yq$ and appears in output 
$q$ (Fig.~\ref{gateinputs}b). 

When plasmodia are inoculated in both inputs $x$ and $y$ they appear in both outputs $p$ and $q$
(Fig.~\ref{gateinputs}c--e). In some cases plasmodia originated in different inputs avoid each other (Fig.~\ref{gateinputs}cd)
and thus head towards different outputs. In other cases the plasmodia merge in a single plasmodium but nevertheless branch towards
different outputs (Fig.~\ref{gateinputs}e). There are no strict rules on repelling and merging and often initial repelling between two plasmodia can be followed by merging (Fig.~\ref{gateinputs}fg).

\begin{figure}
\centering
\subfigure[]{
\begin{tabular}{cc|ccc}
$x$ & $y$ & $p$ & $q$ & frequency \\ \hline
0 & 0 & 0 & 0 & 0\\ \hline
0 & 1 & 0 & 0 & 0 \\
  &   & 0 & 1 & $\frac{17}{22}$ \\
  &   & 1 & 0 & $\frac{2}{22}$ \\
  &   & 1 & 1 & $\frac{3}{22}$ \\ \hline
1 & 0 & 0 & 0 & 0 \\
  &   & 0 & 1 & $\frac{9}{13}$ \\
  &   & 1 & 0 & $\frac{3}{13}$ \\
  &   & 1 & 1 & $\frac{1}{13}$ \\ \hline
1 & 1 & 0 & 0 & 0 \\
  &   & 0 & 1 & $\frac{1}{7}$ \\
  &   & 1 & 0 & $0$ \\
  &   & 1 & 1 & $\frac{6}{7}$ \\ \hline  
\end{tabular}
}
\subfigure[]{
\begin{tabular}{cc|ccc}
$x$ & $y$ & $p$ & $q$ & frequency \\ \hline
0 & 0 & 0 & 0 & 0\\ \hline
0 & 1 & 0 & 0 & 0 \\
  &   & 0 & 1 & $\frac{4}{29}$ \\
  &   & 1 & 0 & $\frac{21}{29}$ \\
  &   & 1 & 1 & $\frac{4}{29}$ \\ \hline
1 & 0 & 0 & 0 & 0 \\
  &   & 0 & 1 & $\frac{16}{27}$ \\
  &   & 1 & 0 & $\frac{5}{27}$ \\
  &   & 1 & 1 & $\frac{6}{27}$ \\ \hline
1 & 1 & 0 & 0 & 0 \\
  &   & 0 & 1 & $\frac{13}{21}$ \\
  &   & 1 & 0 & $\frac{4}{21}$ \\
  &   & 1 & 1 & $\frac{5}{21}$ \\ \hline  
\end{tabular}
}
\caption{Experimental data on transformations 
$\langle x, y \rangle \rightarrow \langle p, q \rangle$ implemented by Physarum gates 
(a)~$P_1$ and (b)~$P_2$. Values $x=1$ and $y=1$ are represented by 
plasmodia inoculated in inputs $x$ and $y$, respectively. Values $p=1$ and $q=1$ 
are represented by plasmodia reaching outputs $p$ and $q$, respectively. Frequency 
of each particular scenario $\langle a, b \rangle \rightarrow \langle c, d \rangle$ 
is presented by fraction: denominator is a total number of experiments for 
 $\langle x, y \rangle = \langle a, b \rangle$ and numerator is a number of 
 experiments completed with output tuple $\langle x, y \rangle = \langle c, d \rangle$.  
}
\label{experimentstatistics}
\end{figure}

Typically for biological substrates, plasmodium of \emph{Physarum polycephalum} is sensitive to 
experimental conditions. The plasmodium sometimes deviates from general scenario when
implementing transformation $\langle x, y \rangle \rightarrow \langle p, q \rangle$. 
Results of experiments with Physarum gate $P_1$ are shown in Fig.~\ref{experimentstatistics}a.

Plasmodia do not cancel each other at once. Therefore if at least one of the inputs is '1' we expect to 
see '1' at one of the outputs. Input scenario $\langle 1, 1 \rangle$ is straightforward: in six out of 
seven experiments plasmodia appear on both outputs. Thus we  obtain transformation 
$\langle 1, 1 \rangle \rightarrow \langle 1, 1 \rangle$ (Fig.~\ref{experimentstatistics}a). 

Plasmodium inoculated in input $y$ (while input $x$ is empty) will appear in output $q$ in 17 out of 22 experiments. Thus the transformation $\langle 0, 1 \rangle \rightarrow \langle 0, 1 \rangle$ is realized by plasmodium in over 70\%. 

Input combination $\langle 1, 0 \rangle$ gives us less stable results: in nine our of 13 experiments (69\%) the plasmodium reaches
output $q$. The plasmodium enters output $p$ in three of 13 experiments, and the plasmodium branches inside both 
outputs in one experiment. Nevertheless the transformation $\langle 1, 0 \rangle \rightarrow \langle 0, 1 \rangle$ is realized
by plasmodium in well over half of experiments (Fig.~\ref{experimentstatistics}a).

\begin{finding}
Plasmodium of \emph{Physarum polycephalum} implements two-input two-output Boolean gate  $\langle  x, y \rangle \rightarrow \langle  xy, x+y \rangle$ with reliability exceeding 69\%.
\end{finding}

\begin{figure}
\centering
\subfigure[]{\includegraphics[width=0.3\textwidth]{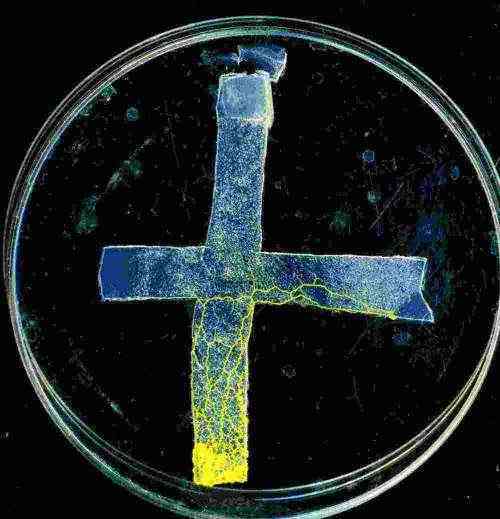}}
\subfigure[]{\includegraphics[width=0.3\textwidth]{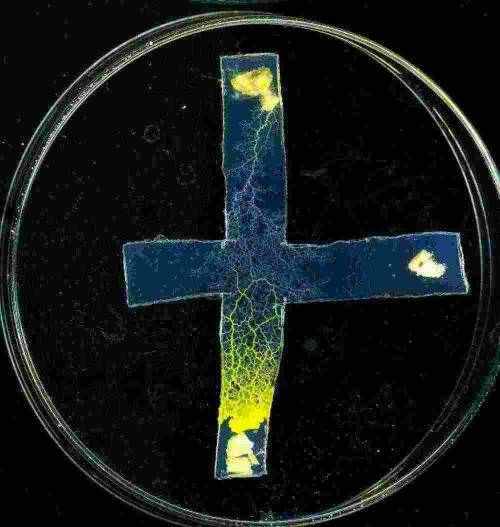}}
\subfigure[]{\includegraphics[width=0.3\textwidth]{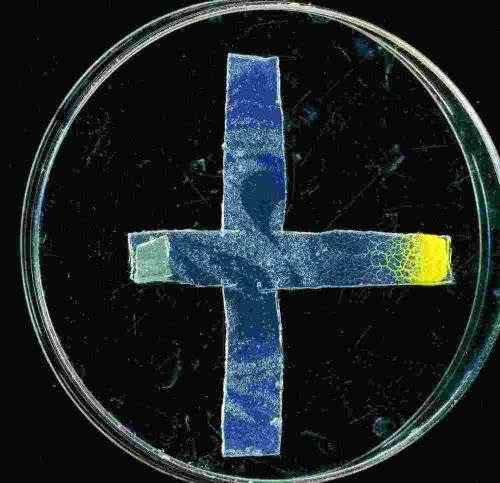}}
\subfigure[]{\includegraphics[width=0.3\textwidth]{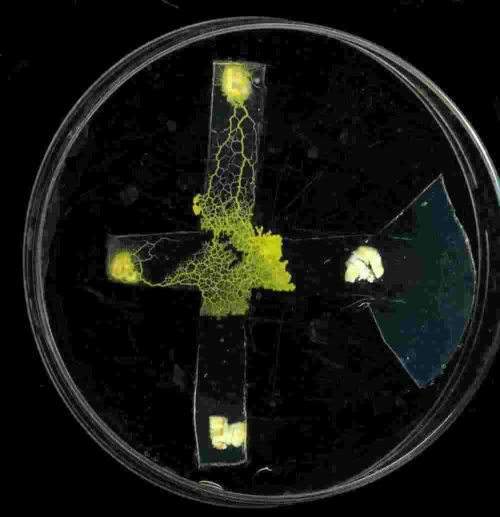}}
\subfigure[]{\includegraphics[width=0.3\textwidth]{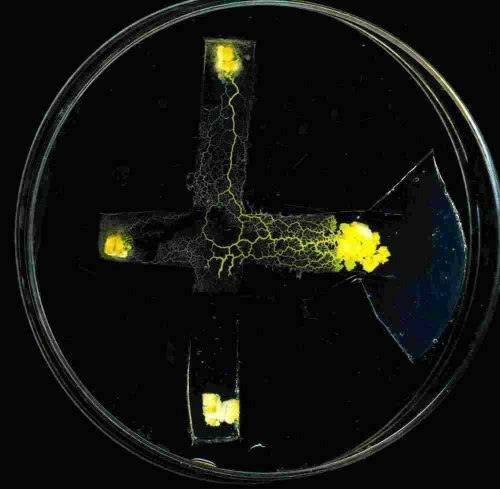}}
\subfigure[]{\includegraphics[width=0.3\textwidth]{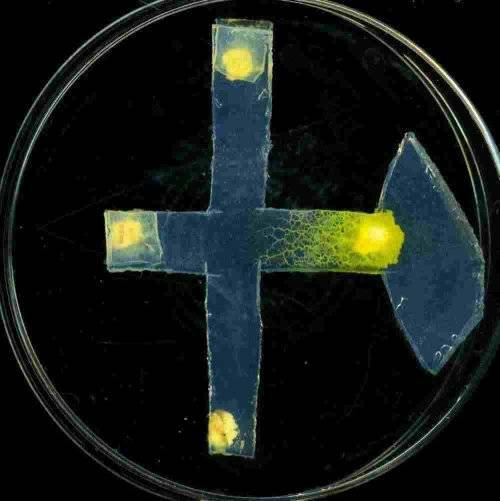}}
\subfigure[]{\includegraphics[width=0.3\textwidth]{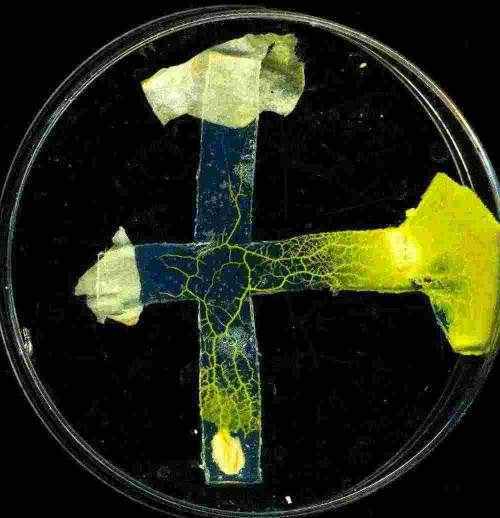}}
\subfigure[]{\includegraphics[width=0.3\textwidth]{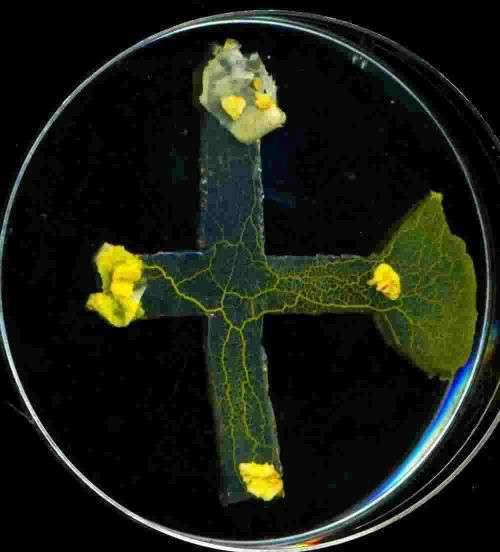}}
\caption{Experimental examples of transformation $\langle x, y \rangle \rightarrow \langle p, q \rangle$
implemented by Physarum gate $P_2$:
(a)~$\langle 0, 1 \rangle \rightarrow \langle 1, 0 \rangle$, plasmodium is inoculated in input $y$, no oat flakes present;
(b)~ $\langle 0, 1 \rangle \rightarrow \langle 1, 0 \rangle$, oat flakes are placed in both outputs;
(c)~$\langle 1, 0 \rangle \rightarrow \langle 0, 1 \rangle$, no oat flakes present;
(d)--(e)~two snapshots of transformation $\langle 1, 1 \rangle \rightarrow \langle 0, 1 \rangle$, taken with 11~h interval,
oat flakes are placed in both outputs;
(f)~$\langle 1, 1 \rangle \rightarrow \langle 0, 1 \rangle$, oat flakes are placed in both outputs;
(g)--(h)~transformations $\langle 1, 1 \rangle \rightarrow \langle 0, 1 \rangle$ are less pronounced than in previous examples, however we see that output $q$ is more extensively occupied by plasmodium than output $p$. }
\label{notgate}
\end{figure}

Experimental snapshots of plasmodium propagating in the gate $P_2$ are shown in Fig.~\ref{notgate}. Taken input $x$ is empty, plasmodium placed in input $y$ usually (see statistics in Fig.~\ref{experimentstatistics}) propagates directly 
towards output $q$ (Fig.~\ref{notgate}ab).
Plasmodium inoculated in input $x$ (when input $y$ is empty) travels directly towards output $p$ (Fig.~\ref{notgate}c). Thus transformations $\langle 0, 1 \rangle \rightarrow \langle 1, 0 \rangle$ and 
$\langle 1, 0 \rangle \rightarrow \langle 0, 1 \rangle$ are implemented.

The gate's structure is asymmetric,  $x$-channel is shorter than $y$-channel. Therefore the plasmodium placed in input $x$ usually passes the junction by the time plasmodium originated in input $y$ arrives at the junction (Fig.~\ref{notgate}d--f).
The $y$-plasmodium merges with $x$-plasmodium and they both propagate towards output $q$. Extension of gel substrate after output $q$ does usually facilitate implementation of the transformation $\langle 1, 1 \rangle \rightarrow \langle 0, 1 \rangle$ (Fig.~\ref{notgate}d--h).  

Frequencies of various input-output transformations occurred in experiments are shown in Fig.~\ref{experimentstatistics}b. 
Plasmodium inoculated in input $y$ will reach only output $p$ in 21 out of 29 experiments. Transformation  
$\langle 1, 0 \rangle \rightarrow \langle 0, 1 \rangle$ takes place in 16 out of 27 experiments. Transformation
$\langle 1, 1 \rangle \rightarrow \langle 0, 1 \rangle$ occurs in 13 out of 21 experiments.

\begin{finding}
Plasmodium of \emph{Physarum polycephalum} implements two-input two-output gate  
$\langle  x, y \rangle \rightarrow \langle  x, \overline{x}y \rangle$ with 
reliability exceeding 59\%.
\end{finding}

\section{Simulation of Physarum gates}
\label{simulation}

Experimental Findings 1 and 2 are confirmed in numerical simulation using two-variable Oregonator
model of a sub-excitable medium (see Sect.~\ref{methods}). To represent value '1' in input channel 
$x$ or channel $y$ 
 we generate an excitation near entrance of the channel. Two wave-fragments are formed. One travels
 outside the gate, another travels towards outputs.

\begin{figure}
\centering
\subfigure[]{\fbox{\includegraphics[width=0.2\textwidth]{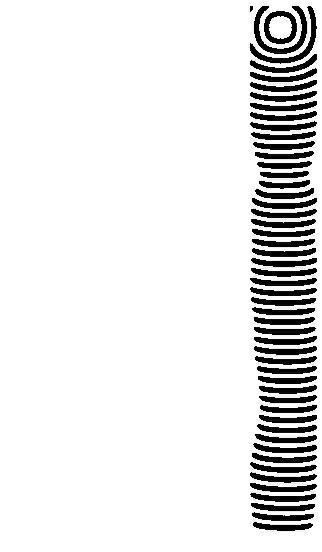}}}
\subfigure[]{\fbox{\includegraphics[width=0.2\textwidth]{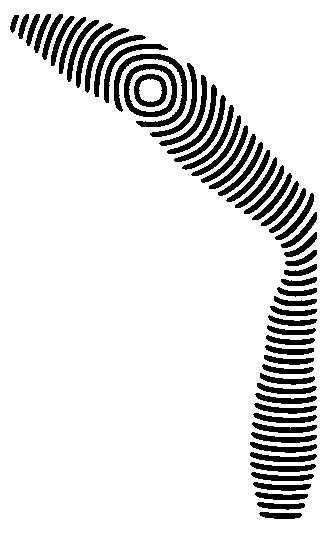}}}
\subfigure[]{\fbox{\includegraphics[width=0.2\textwidth]{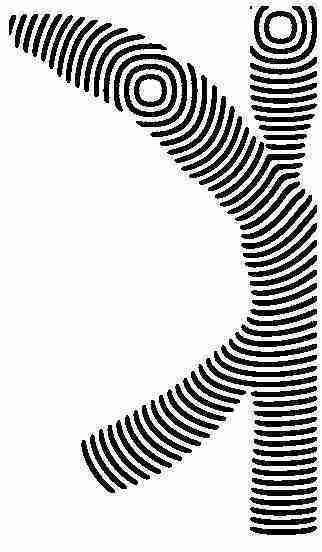}}}
\caption{Time lapsed images of localized excitations, wave-fragments, traveling in  
channels of gate $P_1$ filled with excitable medium. Dynamics of the excitable medium
gate $P_1$ is shown during implementation of transformations 
(a)~$\langle 0, 1 \rangle \rightarrow \langle 0, 1 \rangle$,
(b)~$\langle 1, 0 \rangle \rightarrow \langle 0, 1 \rangle$,
(c)~$\langle 1, 1 \rangle \rightarrow \langle 1, 1 \rangle$. 
The transformation $\langle 1, 1 \rangle \rightarrow \langle 1, 1 \rangle$  is
simulated for initial excitations in channels $x$ and $y$ positioned at
equal distance from their meeting point. }
\label{simulationfig}
\end{figure}

\begin{figure}
\centering
\subfigure[]{\fbox{\includegraphics[width=0.29\textwidth]{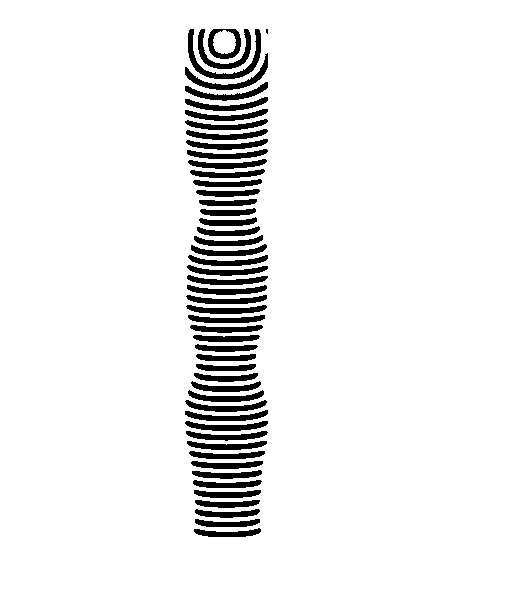}}}
\subfigure[]{\fbox{\includegraphics[width=0.29\textwidth]{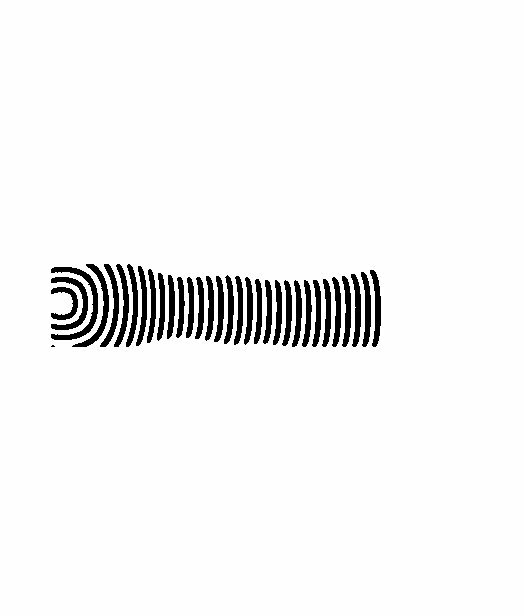}}}
\subfigure[]{\fbox{\includegraphics[width=0.29\textwidth]{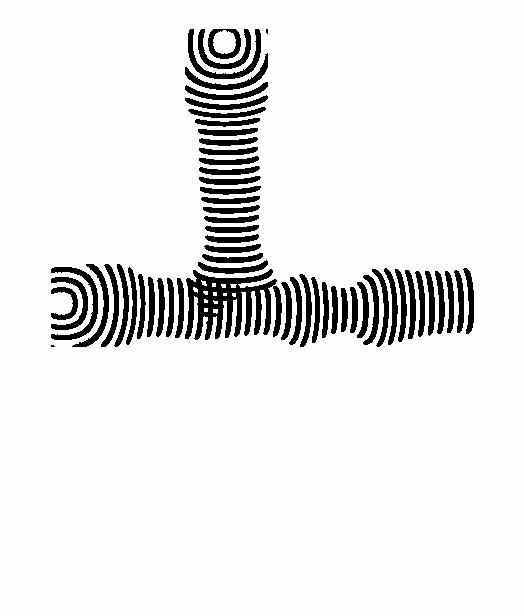}}}
\caption{Time lapsed images of localized excitations, wave-fragments, traveling in  
channels of gate $P_2$ filled with excitable medium. Dynamics of the excitable medium
gate $P_2$ is shown during implementation of transformations 
(a)~$\langle 0, 1 \rangle \rightarrow \langle 1, 0 \rangle$,
(b)~$\langle 1, 0 \rangle \rightarrow \langle 0, 1 \rangle$,
(c)~$\langle 1, 1 \rangle \rightarrow \langle 0, 1 \rangle$. }
\label{simulationfig2}
\end{figure}

Wave-fragments in sub-excitable media are notably unstable, they do keep their shape only for a short period of time. Then the fragments either shrink and annihilate or expand unlimitedly. During simulation of gates $P_1$ and $P_2$ we manually adjusted parameter 
$\epsilon$ (see Sect.~\ref{methods}) to keep wave-fragments from collapsing and expanding. Figures~\ref{epsilon} and \ref{notepsilon} illustrate dynamics of $\epsilon$ during simulations of gates $P_1$ (Fig.~\ref{simulationfig}) and 
$P_2$ (Fig.~\ref{simulationfig2}).

Let us look at the time lapsed  images of wave-fragments propagating in gate $P_1$ (Fig.~\ref{simulationfig}) and gate 
$P_2$ (Fig.~\ref{simulationfig2}). Scenarios where only one input is excited are straightforward. When $y=1$ in gate $P_1$
is initialized the wave-fragment propagates southward along the channel $yq$ (Fig.~\ref{simulationfig}a). Wave-fragment 
initiated in input $x$ ($x=1$) of gate $P_1$ propagates along input channel $x$ and collides to 
the boundary of channel $yq$ (Fig.~\ref{simulationfig}b). The wave-fragment recovers after collision to the boundary, 
travels along the channel $yq$ and appears in the output $q$ (Fig.~\ref{simulationfig}b). Wave-fragments behave
similarly in situations of input tuples $\langle 1, 0 \rangle$ and $\langle 0, 1 \rangle$ in gate $P_2$. They propagate straight along their original input channel and reach outputs opposite to their entry points 
(Fig.~\ref{simulationfig2}ab).  

In input scenarios $\langle 0, 1 \rangle$ and $\langle 1, 0 \rangle$ size of propagating wave-fragment was not enough for 
the fragment to branch into output channel $p$ of gate $P_1$ (Fig.~\ref{simulationfig}ab). When two wave-fragments are initiated,  $x=1$ and $y=1$, they collide at the junction of input channels $x$ and $y$. The wave-fragments merge into a single larger-wave fragment.  This new fragment propagates towards output $q$ and also expands into output channel $p$ (Fig.~\ref{simulationfig}c).

In gate $P_2$, inputs $x=1$ and $y=1$, wave-fragment originated in $x$-input arrives at the junction between channel $xp$ and 
$yq$ a bit earlier then wave-fragment originated in $y$-input. Therefore $y$-wave collides into refractory tail of $x$-wave. The 
$y$-wave annihilates (Fig.~\ref{simulationfig2}c). Such development is phenomenologically identical to 
Physarum gate $P_2$ (Fig.~\ref{notgate}de) with the only difference that plasmodium `wave' originated in input $y$ is not annihilated but merges with plasmodium originated in input $x$.

\section{Simulated one-bit half-adder}
\label{adder}

\begin{figure}
\centering
\includegraphics[width=0.6\textwidth]{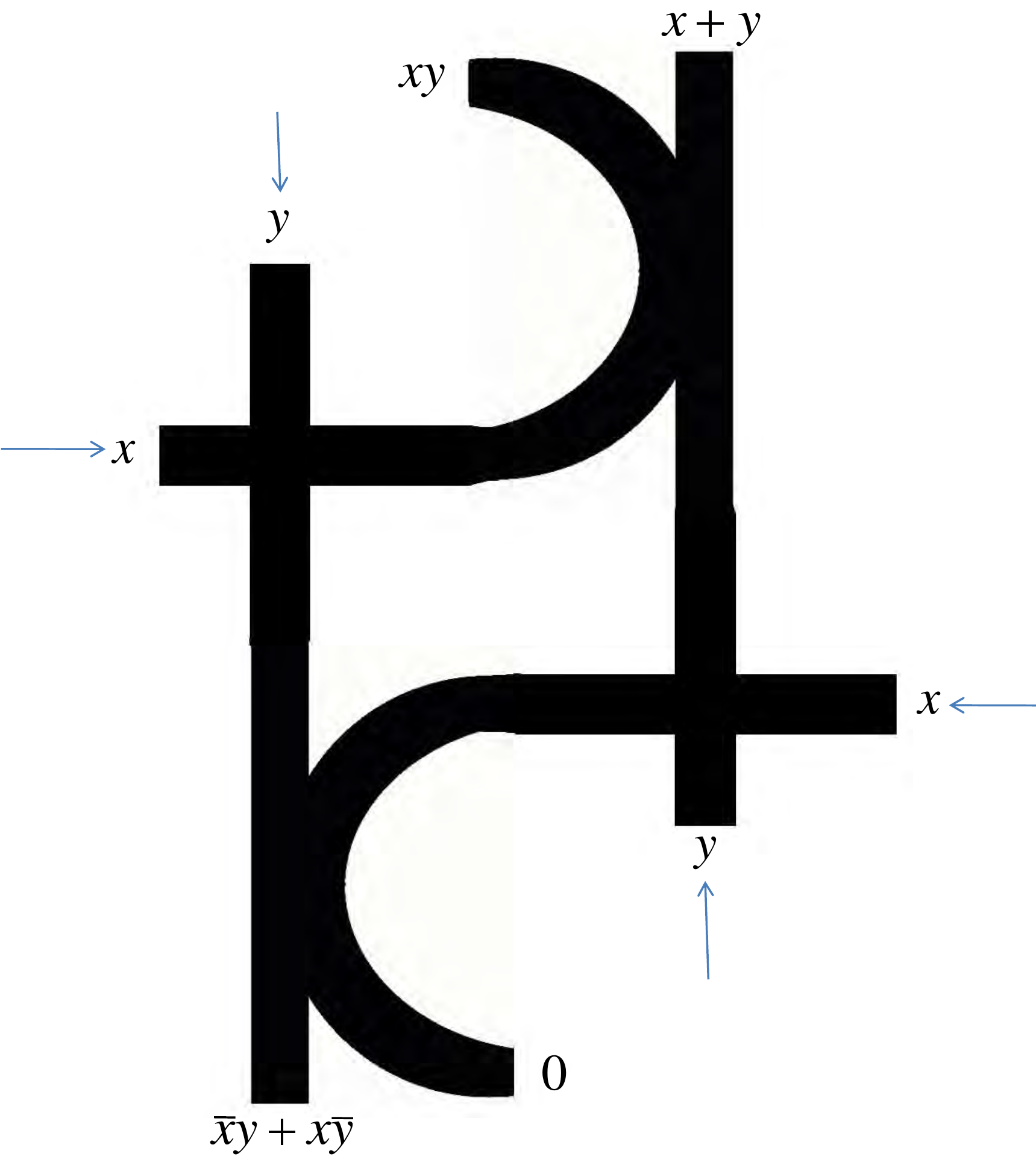}
\caption{Scheme of one-bit half-adder made of gates $P_1$ and $P_2$. Inputs are indicated by arrows. Outputs 
$\overline{x}y + x\overline{y}$ and $xy$ are sum and carry values computed by the adder. Outputs $0$ and 
$x+y$ are byproducts.}
\label{adderscheme}
\end{figure}

One-bit half-adder is a logical circuit which takes two inputs $x$ and $y$ and produces two outputs: 
sum $\overline{x}y + x\overline{y}$ and carry $xy$. To construct a one-bit half-adder with Physarum gates
we need two copies of gate $P_1$ (Fig.~\ref{gatescheme}a) and two copies of gate $P_2$ (Fig.~\ref{gatescheme}b).
Cascading the gates into the adder is shown in Fig.~\ref{adderscheme}. Signals $x$ and $y$ are inputted in $P_2$ gates. 
Outputs of $P_2$ gates are connected to inputs of $P_1$ gates.

\begin{figure}
\centering
\subfigure[]{\fbox{\includegraphics[width=0.45\textwidth]{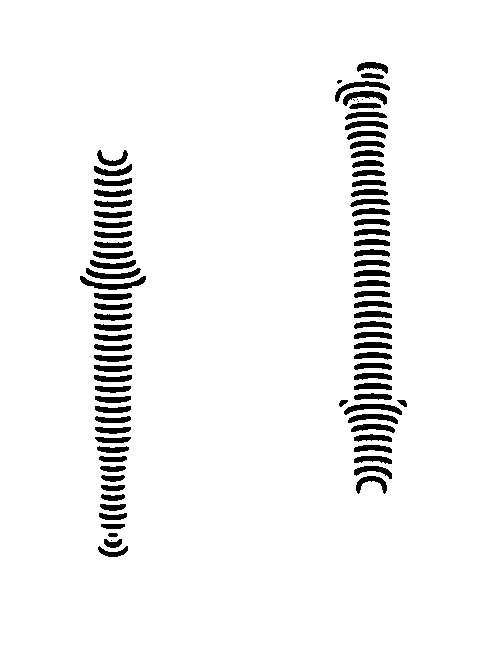}}}
\subfigure[]{\fbox{\includegraphics[width=0.45\textwidth]{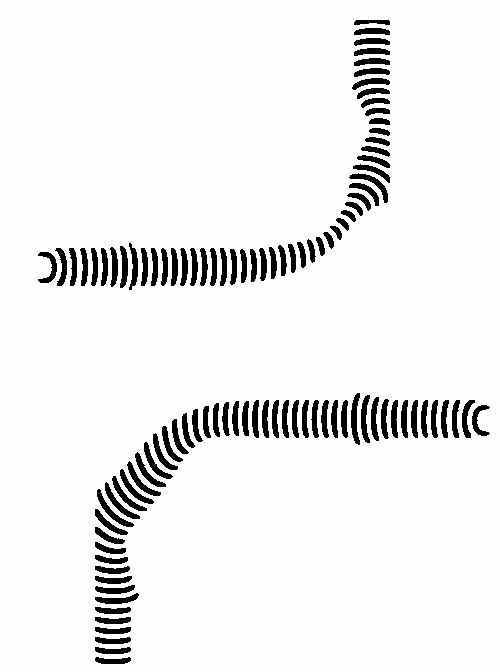}}}
\subfigure[]{\fbox{\includegraphics[width=0.45\textwidth]{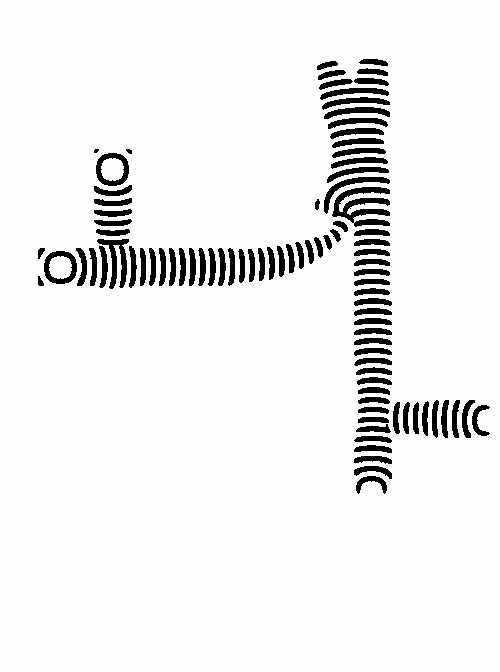}}} \hspace{0.5cm}
\caption{Time lapsed images of localized excitations, wave-fragments, traveling in  
channels of one-bit half-adder filled with excitable medium. Dynamics of excitations
is shown for input values 
(a)~$x=0$ and $y=1$, 
(b)~$x=1$ and $y=0$, 
(c)~$x=1$ and $y=1$.}
\label{lapsedadder}
\end{figure}

We did not manage to realize one-bit half-adder in experiments with living plasmodium because the plasmodium behaved differently 
in the assembly of the gates than in isolated gates. Simulation of the adder using Oregonator model was successful (Fig.~\ref{lapsedadder}). To simulate inputs $x=0$ and $y=1$ we initiate wave-fragments at the beginning of channels, marked $y$ and arrow in Fig.~\ref{adderscheme}. The wave-fragments propagate along their channels. The waves do not branch at the junctions with other channels because we keep the wave-fragments localized by varying parameter $\epsilon$ (Fig.~\ref{lapsedadder}a).

For input values $x=1$ and $y=0$ wave-fragments are originated in sites marked $x$ and arrow in Fig.~\ref{adderscheme}. The wave-fragment started in left $x$-input channel propagates towards $x+y$-output of the adder. The wave-fragment
originated in right $x$-input channel travels towards $\overline{x}y + x \overline{y}$ (Fig.~\ref{lapsedadder}b).

When both inputs are activated, $x=1$ and $y=1$, wave-fragment originated in left $y$-input channel is 
blocked by refractory tail of wave-fragment originated in left $x$-input channels. The wave-fragment traveling in right $x$-input channel is blocked by tail of wave-fragment traveling in right $y$-input channel. The wave-fragments representing $x=1$ and $y=1$ enter top-right gate $P_1$ and emerge at its outputs $xy$ and $x+y$ (Fig.~\ref{lapsedadder}b). Thus functionality of the circuit Fig.~\ref{lapsedadder} is demonstrated.

\section{Discussion}
\label{discussion}

We established experimentally and in numerical simulations that plasmodium of \emph{Physarum polycephalum} realizes basic logical operations on a geometrically-constrained non-nutrient substrate.  We designed two types of Boolean logic gates, both gates have two inputs and two outputs. The gates implement transformations  $\langle  x, y \rangle \rightarrow \langle  xy, x+y \rangle$ and 
 $\langle  x, y \rangle \rightarrow \langle  x, \overline{x}y \rangle$. We shown how the Physarum gates can be assembled into a
 one-bit half-adder. Functionality of the adder is illustrated using two-variable Oregonator model.

\begin{figure}
\centering
\subfigure[]{\includegraphics[width=0.345\textwidth]{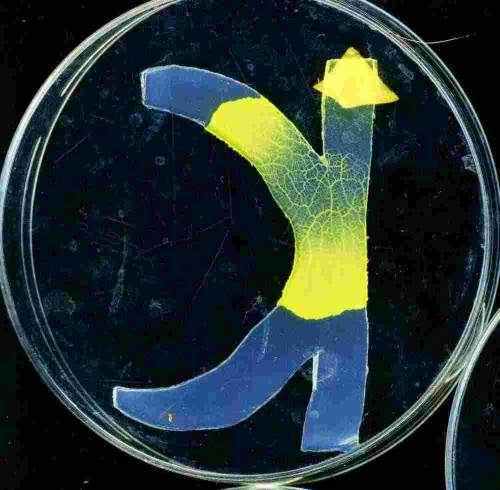}}
\subfigure[]{\includegraphics[width=0.31\textwidth]{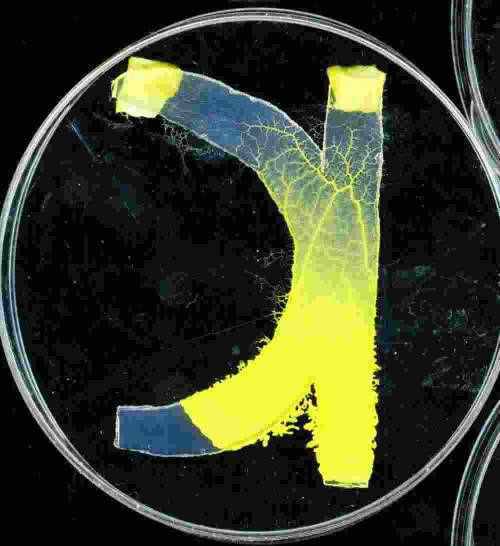}}
\subfigure[]{\includegraphics[width=0.33\textwidth]{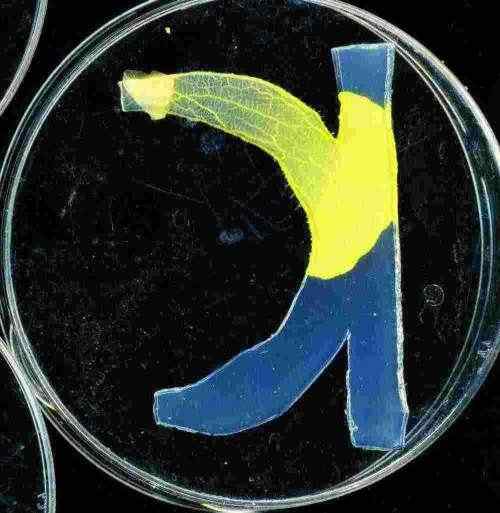}} 
\subfigure[]{\includegraphics[width=0.2\textwidth]{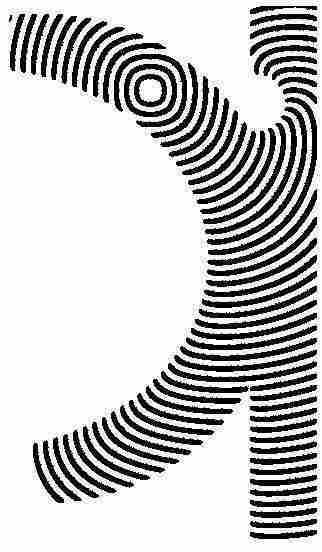}}
\caption{Snapshots of plasmodium propagating in gate $P_1$ on 2\% corn meal agar (a--c) and numerical simulation (d):
(a)~inputs $x=0$ and $y==1$, see
the gate's scheme in Fig.~\ref{gatescheme}, (b)~inputs $x=1$ and $y=1$, (c)~inputs $x=1$ and $y=0$, 
(d)~time lapsed images of excitation wave propagating in gate $P_1$ for inputs $x=1$ and $y=0$, simulation of 
case (c).}
\label{gatesnutrients}
\end{figure}

Our designs are based on the interactions between traveling localizations: plasmodium localizations propagating on a non-nutrient substrate and wave-fragments propagating in a sub-excitable medium. Similarities between the plasmodium localizations and wave-fragments are discussed in details~\cite{adamatzky_physarum_bz}. We stress that things go absolutely differently 
on a nutrient-rich substrate (corn meal agar) and fully excitable chemical medium. Plasmodium inoculated in any point of 
the nutrient-agar gel gate propagates in all channels (Fig.~\ref{gatesnutrients}a--c). An excitation wave initiated at any point of excitable medium gate spreads all over the gate (Fig.~\ref{gatesnutrients}d). We can conclude therefore that it is impossible to implement logical functions with plasmodium of \emph{Physarum polycephalum} on a nutrient-rich substrate. 

Reliability of experimental Physarum gate is quite low: 69\% for gate $P_1$ and 59\% for gate $P_2$. This is because
behavior of plasmodium is determined by too many environmental factors --- thickness of substrate, humidity, diffusion of chemo-attractants in the substrate and in the surrounding air volume, and physiological  state of plasmodium during each particular experiment. Increasing reliability of Physarum gates might be a scope of further studies.

\newpage

\section*{Appendix A}

\begin{figure}
\centering
\subfigure[]{\includegraphics[height=0.27\textheight]{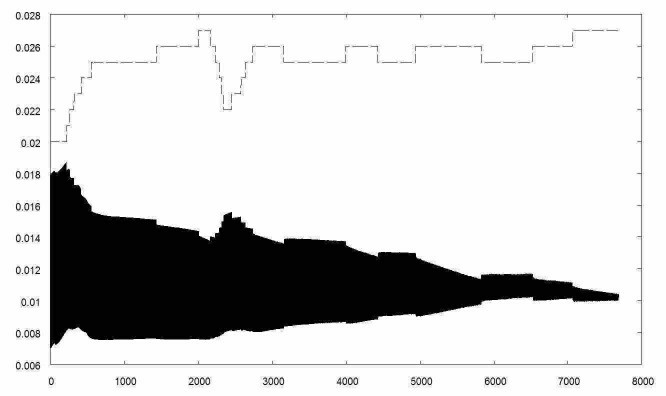}}
\subfigure[]{\includegraphics[height=0.27\textheight]{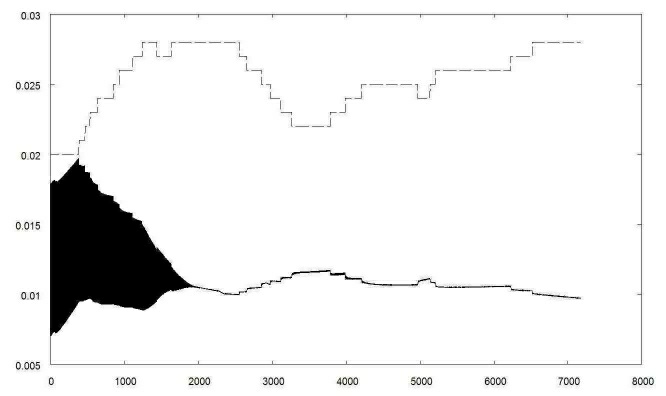}}
\subfigure[]{\includegraphics[height=0.27\textheight]{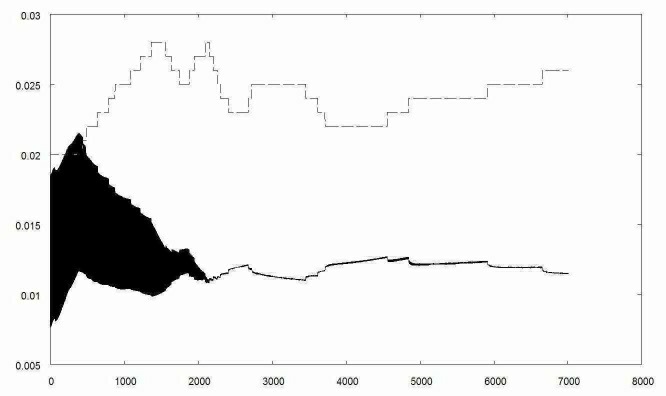}}
\caption{Dynamics of parameter $\epsilon$, dotted line, and 
activity $\alpha$, solid line, during operation cycle of gate $P_1$ for 
input tuples
(a)~$\langle 0, 1 \rangle$ (see Fig.~\ref{simulationfig}a), 
(b)~$\langle 1, 0 \rangle$ (see Fig.~\ref{simulationfig}b),
(c)~$\langle 1, 1 \rangle$ (see Fig.~\ref{simulationfig}c).}
\label{epsilon}
\end{figure}

\begin{figure}
\centering
\subfigure[]{\includegraphics[height=0.27\textheight]{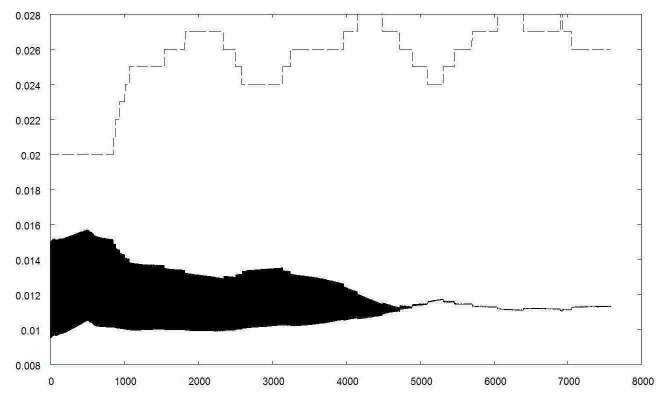}}
\subfigure[]{\includegraphics[height=0.27\textheight]{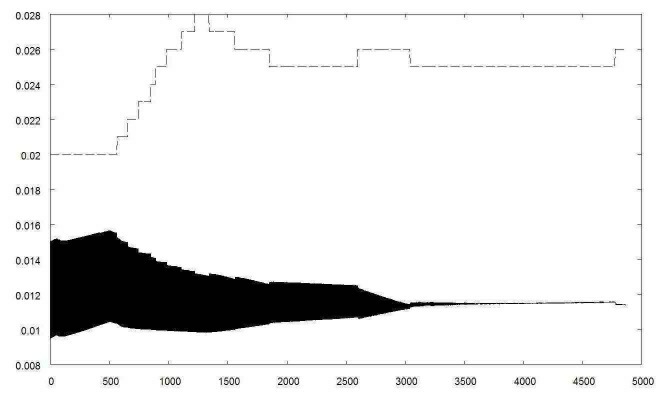}}
\subfigure[]{\includegraphics[height=0.27\textheight]{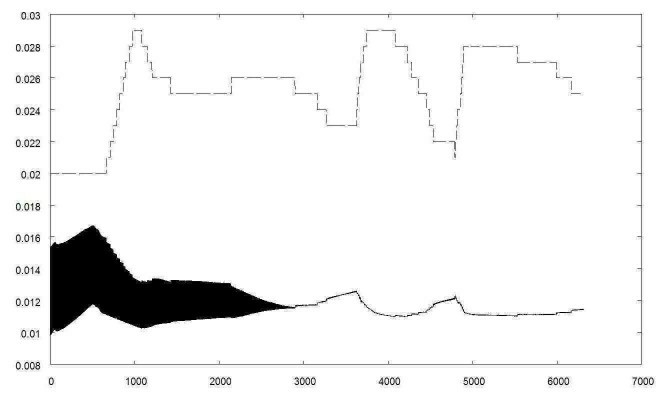}}
\caption{Dynamics of parameter $\epsilon$, dotted line, and 
activity $\alpha$, solid line, during operation cycle of gate $P_2$ for 
input tuples
(a)~$\langle 0, 1 \rangle$ (see Fig.~\ref{simulationfig2}a), 
(b)~$\langle 1, 0 \rangle$ (see Fig.~\ref{simulationfig2}b),
(c)~$\langle 1, 1 \rangle$ (see Fig.~\ref{simulationfig2}c).}
\label{notepsilon}
\end{figure}

Figures~\ref{epsilon} and~\ref{notepsilon} show how  parameter $\epsilon$ (see description of
the model in Sect.~\ref{methods}) is changed during simulation of gates $P_1$ and $P_2$ in a 
sub-excitable medium. We also show dynamics of the medium's activity. At each step of simulation we calculate 
activity $\alpha^t = |\mathbf{L}|^{-1} \cdot \sum_{x \in \mathbf{L}} u^t_x$  as a sum of values $u$ for each node of the grid $\mathbf L$ normalized by total number of nodes $|\mathbf{L}|$. 

\end{document}